\documentclass[final,3p,times]{article}
\usepackage{amssymb}
\usepackage{amsmath}
\usepackage{amsfonts}
\usepackage{graphicx}
\usepackage{authblk}
 \title{On inverse product cannibalisation: \\
 a new Lotka-Volterra model for asymmetric competition in the ICTs}

\author[1]{Mariangela Guidolin}
\author[1]{Renato Guseo}
\affil[1]{Department of Statistical Sciences, University of Padua}

\begin{document}

\maketitle

\begin{abstract}
Product cannibalisation is a well-known phenomenon in marketing and technological research and describes the case when a new product steals sales from another product under the same brand. A very special case of cannibalisation may occur when the older product react to the competitive strength of the newer one, absorbing the corresponding market shares. Given its special character, we call this phenomenon \textit{inverse product cannibalisation}. We suppose that a case of inverse cannibalisation is observed between two products of Apple Inc., the iPhone and the more recent iPad, and the first has been able to succeed at the expense of the second. To explore this hypothesis, within a diffusion of innovations perspective, we propose a modified Lotka-Volterra model for mean trajectories in asymmetric competition, allowing us to test the presence and the extent of the inverse cannibalisation phenomenon. A SARMAX refinement integrates the short term predictions with seasonal and autodependent components. 
A non-dimensional representation of the proposed model shows that the penetration of the second technology has been beneficial for the first, both in terms of market size and life cycle length. 
\end{abstract}


%\end{frontmatter}


\section{Introduction}
Sales dynamics may be considered as complex systems affected by exogenous and endogenous shocks. The distinction between exogenous and endogenous has been highlighted by \cite{sornette:04}: exogenous peaks in sales occur abruptly and are followed by power law relaxation, while endogenous occur after a power law growth followed by a power law relaxation. 
An analogous perspective has been taken in the innovation diffusion context, \cite{bass:94}. Among all kinds of shocks that may affect sales dynamics, competition between products plays a central role. 
In this paper we consider a special kind of competition, namely cannibalisation, and investigate its effects. 
Product cannibalisation is a well-known phenomenon in marketing and technology research and refers to the situation where two or more products from the same brand compete and take away market shares from each other \cite{thota:11}. In \cite{copulsky:76} product cannibalisation is defined as `the extent to which one product's sales are at the expense of other products offered by the same firm'. Although the effect of cannibalisation may appear negative, it may eventually be beneficial, bringing in new consumers and growing the overall market size. In fact, cannibalisation is considered a common strategy in some industrial sectors such as computer hardware and software, financial services, airline services, automotive, food, and pharmaceuticals \cite{thota:11}. This strategy is typically indicated as \textit{planned self-cannibalisation}, and it is successful when the underlying technology continuously advances. 

So far, the marketing and technology literature has focused on how new products  cannibalise existing ones \cite{bordley:17}, \cite{davis:06} and \cite{vanheerde:10}. 
As highlighted by \cite{vanheerde:10}, to evaluate the success of a new product, managers need a method to verify not only how much demand it generates, but also to what extent this demand comes at the expense of their other products. Thus, an assessment of the extent of cannibalisation is crucial for understanding whether the introduction can be considered a success for the firm as a whole. 

The phenomenon of cannibalisation is stronger when firms are active in several commercial categories and introduce pioneering innovations, \cite{vanheerde:10}. 
Typical examples of firms affected by this phenomenon are Unilever, Hewlett-Packard, Procter and Gamble, and Apple. 
Indeed, an often cited case of cannibalisation in the ICT sector, see \cite{bordley:17} and \cite{vanheerde:10}, is that between Apple's iPod and iPhone: Apple's first quarter 2013 earnings report  acknowledged iPhones cannibalising iPods. This happened because the iPhone occupied two commercial categories: portable media players and mobile phones. When introducing the iPhone, Steve Jobs said `if anyone is going to cannibalise us, I want it to be us. I don't want it to be a competitor'. 
\begin{figure}
\begin{center}
\includegraphics[scale=0.14]{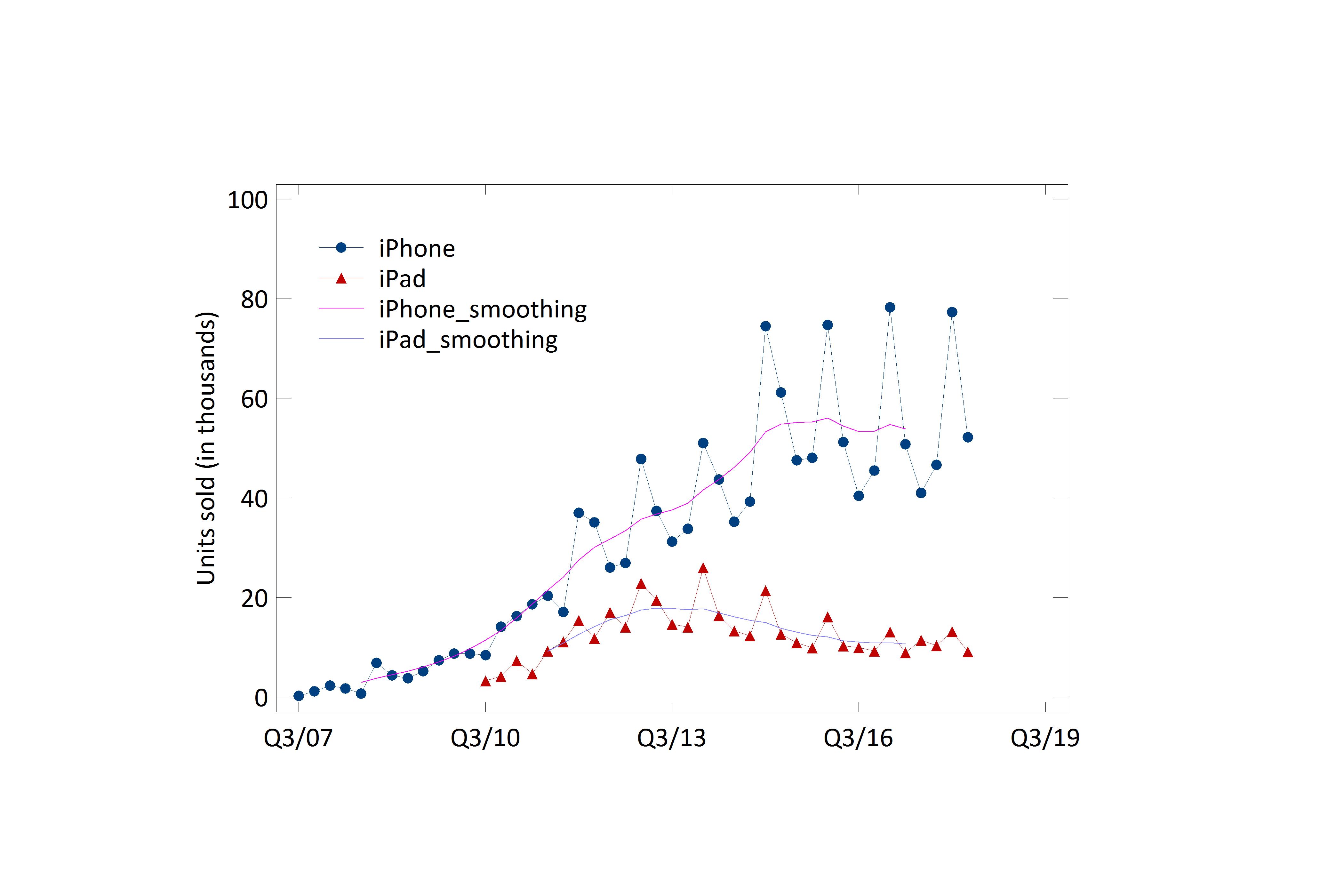}
\vspace{-22mm}
\caption{Quarterly actual and smoothed units of iPhone and iPad sold. Smoothing performed through a five terms moving average. (data source: Apple Inc).}
\label{fig1}
\end{center}
\end{figure}
Although self cannibalisation appears a common strategy for this brand, when introducing a new product, an intriguing question arises when inspecting Fig. \ref{fig1}, which displays actual and smoothed (deseasonalised) quarterly sales of the iPhone and iPad at the world level. Through a five terms smoothing we may better capture the mean behaviour of the two series, observing that the iPhone has been experiencing a growing trend, while the more recent iPad has apparently overtaken the maximum peak of its finite life cycle. Moreover, the market share of the iPad appears significantly smaller than that of the iPhone. 
This may seem surprising, because one could expect the iPad to have greater success. Given this, one could make the hypothesis that the decline in sales of the iPad and the parallel success of the iPhone are tied by competitive dynamics. Even though the two technologies are not substitutes or subsequent generations of each other, they share many features and functionalities that may imply a degree of interchangeability; if the traditional telephony is a specified service of the iPhone, many other features are in common, such as Internet, browsing services, GPS services (maps, dynamic positioning), document generation, photographs, video and archiving, written and oral messages with possible attachments, email, Internet telephony, multi-channel contact directory and music reproduction.
Stimulated by the fact that Apple's marketing strategy encourages cannibalisation among its products and the company itself recognised the phenomenon, we provide a statistical model in order to test whether and to which extent the competitive strength of the iPhone gave rise to a very special case of cannibalisation, that we may call \textit{inverse product cannibalisation}, where it is the first product (iPhone) that takes away market share of the second (iPad). 
To do so, we adopt an innovation diffusion perspective generalising the basic univariate Bass model \cite{bass:69} and propose a bivariate life cycle model for competing products, which we call \textit{Lotka-Volterra model with asymmetric competition}, LVac. This model is derived as a special case of the Lotka-Volterra with churn model, LVch, proposed by \cite{guidolin:15}.

Once the significant presence of inverse product cannibalisation has been tested and selected, an important question remains open: what are the consequences of cannibalisation on a product life cycle? 
Because strategic and operational decision making  may be guided by a correct evaluation of its impact \cite{srinivasan:05}, what are the benefits of this, as a brand strategy? To answer this question, we study a non-dimensional representation of the proposed LVac model, which allows for a reduction of the involved parameters and helps in understanding how inverse product cannibalisation modifies the temporal dynamics of an incumbent product's life cycle.\\

The paper is structured as follows: in Section \ref{sect:1}, some key points of finite life cycle models under competition are discussed. In Section \ref{sect:2}, we derive the LVac model with asymmetric competition as a suitable reduction of the more general Lotka-Volterra with churn model, LVch, proposed in \cite{guidolin:15}, and we illustrate some aspects concerning statistical inference and model selection for a fine tuning. In Section \ref{sect:4}, we analyse the case of competition between Apple iPhone and iPad and test our hypothesis on the existence of an inverse cannibalisation effect. Specifically, we show the better performance of the proposed LVac model with that of two univariate standard Bass models. Based on this, in subsection \ref{subsect:2} we perform an out-of-sample forecasting with a two stage procedure: out-of-sample prediction of mean trajectory and SARMAX refinement (see for instance \cite{billings:13}), in order to capture the seasonality and autodependency not modelled with the LVac.  In Section \ref{sect:5}, we propose a non-dimensional representation of the LVac model employed in Section \ref{sect:4}, which allows to better understand what the effect is of the inverse product cannibalisation. Section \ref{sect:6} is devoted to a summary of the results, a discussion and the concluding remarks. 

\section{Competition modelling in a diffusion of innovation context}
\label{sect:1}
One of the key points to consider when studying the penetration of a new product in a market is the presence of competitors. 
Competition across products, technologies or, more generally, scientific paradigms, may be interpreted through complex systems representations obtaining specific asymmetric distributions over time, \cite{jensen:11}.
Competition may alter the life cycle dynamics in terms of penetration speed, time to maximum peak and size of market potential. Concurrent products may act as a barrier to a product's success, but at the same time, their presence may enlarge the size of demand, thus implying a benefit for all market players.
Markets are increasingly becoming complex, and in many commercial sectors the competitive environment counts several actors: in this case, the displacement of older products by a strong competitor may determine the emergence of self--coordination in the collective response of relatively free agents. A signal of this condition is diagnosed by bimodal distributions of the gross income at the opening phase or by bimodal distributions of the total gross income of the total life time of products, \cite{chakrabarti:16b}. 
However, when competitors are few, namely two or three, equations governing their behaviour may be more specific, by defining competition, substitution, cannibalisation or, in some cases, co-existence. 
The models so far produced in the field of new product diffusion, accounting for competition, have usually limited their attention to duopolistic conditions, \cite{peres:10}. This is likely because of the inherent difficulty of managing systems of differential equations, simultaneously describing the mean growth dynamics of each product and their interaction.
More properly, diffusion models for competition have often focused on modelling the interaction between two products by splitting the word-of-mouth (WOM) in two parts: the \textit{within-product} word-of-mouth, which is from the product's specific sales, and the \textit{cross-product} word-of-mouth, which is from competition with the antagonist and may imply either a negative or positive effect. 
Moreover, competition has been considered to be both \textit{synchronic}, which occurs when two products enter the market at the same time, or \textit{diachronic}, which occurs when the first product initially acts as a monopolist and the second enters the market later. 

The literature dealing with technological competition is quite recent but rapidly growing \cite{meade:06}, \cite{peres:10}. Important contributions come from the marketing, operations research and econophysics domains. 
Among others, we note \cite{guseomortarino:10},  \cite{guseomortarino:12}, \cite{guseomortarino:14}, \cite{guseomortarino:15} \cite{krishnan:2000}, \cite{laciana:14} and \cite{savin:05} where competition is modelled through cross-product effects.
In particular, in \cite{guseomortarino:14}, a model called unrestricted \textit{unbalanced competition regime change diachronic model}, UCRCD, where both products share the same market potential and are influenced by within-product and cross-product word-of-mouth, is proposed. In this sense, the residual market is a common target too, which allows completely free competition. This model has been generalised in \cite{guseomortarino:15} by accounting for a time-dependent market potential, $m(t)$. \\
Technological competition has also been modelled with Lotka-Volterra equations: from the first works applying the famous predator-prey relationship described in the Lotka-Volterra models to technological competition \cite{abramson:98}, \cite{balaz:12} and \cite{morris:03}, the contributions have recently been expanded to explore the competitive dynamics occurring between technologies and new ideas  \cite{chakrabarti:16}, \cite{guidolin:15}, \cite{gupta:16}, \cite{tseng:14} and \cite{vitanov:10}.
 
In this context, a special approach for the definition of the residual market has been proposed by \cite{guidolin:15}, extending through a word-of-mouth splitting the standard Lotka-Volterra model, which usually considers interaction based on current input factor, excluding the contribution of the antagonist. This model, characterised by an independent modulation of the residual market of each competitor, takes into account the possibility of a migration of potential adoptions from one competitor to the other, that is, a sort of `churn', and accordingly takes the name Lotka-Volterra with churn model, LVch. This model, along with its possible reductions, will be illustrated in Section \ref{sect:2}. \\

%==================================================================
\section{Lotka-Volterra with churn model, LVch, and its reduction for asymmetric competition, LVac}
\label{sect:2}
%======================================================================
The Lotka-Volterra with churn model, LVch, by \cite{guidolin:15} is a system of differential equations that describe two adoption processes and the related interactions, namely the following:

\begin{eqnarray}\label{1}
z'_1(t)&=& \left[p_{1a}+q_{1a}\frac{z_1(t)}{m_a}\right]\left[m_a-z_1(t)\right], \;\;\;\; t\leqslant{c_2}\nonumber\\
z'_1(t)&=&\left[p_1+\frac{a_1z_1(t)+\alpha_2 b_1z_2(t)}{m_1+\alpha_2 m_2}\right] \left[(m_1-z_1(t))+\alpha_2(m_2-z_2(t))\right]\\
z'_2(t)&=&\left[p_2+\frac{a_2z_2(t)+\alpha_1 b_2z_1(t)}{m_2+\alpha_1 m_1}\right] \left[(m_2-z_2(t))+\alpha_1(m_1-z_1(t))\right]\nonumber.
\end{eqnarray}
In this model, the first equation describes the stand-alone phase ($t \leq c_2$) when the first product acts as a monopolist in the market. We see that the product is assumed to behave according to a standard Bass model \cite{bass:69}, with $z'_1(t)$ and $z_1(t)$ being the rate and cumulative sales, respectively; parameters $m_a$, the market potential, $p_{1a}$, the innovation coefficient due to external information and $q_{1a}$, the imitation coefficient due to word-of-mouth. The second and third equations are defined for $t>c_2$ when the second product has entered the market, and they describe the competition dynamics. 
Each product's rate sales, $z'_i(t), i=1,2$, for $t>c_2$, are proportional to the corresponding residual markets $\left[(m_i-z_i(t))+\alpha_j(m_j-z_j(t))\right], i=1,2, j=1,2$, $i\neq j$, where $m_i$ are the product's specific market potentials under competition, and $z_i(t), i=1,2$, represent the cumulative sales at time $t$. 

As seen here, the residual market is the sum of the product specific one $m_i-z_i(t)$ plus a fraction of the other's, $\alpha_j(m_j-z_j)$. Parameters $\alpha_j,j=1,2$, modulate the size of this second element.
Parameters $p_i,i=1,2$, define the innovative behaviour in adoption, while the word-of-mouth components have a more complex structure, compared with standard LV type models, made of a within-product element $[{a_1z_1(t)}/{(m_1+\alpha_2 m_2)}]$ and a cross-product one, $[{\alpha_2 b_1z_2(t)}/{(m_1+\alpha_2 m_2)}]$ (contribution of the antagonist), for the first competitor and, similarly,  
$[{a_2z_2(t)}/{(m_2+\alpha_1 m_1)}]$ and $[{\alpha_1 b_2z_1(t)}/{(m_2+\alpha_1m_1)}]$ for the second. While parameters $a_1$ and $a_2$ are always positive because they describe the within-product word-of-mouth, which is self-sustaining and therefore positive, parameters $b_1$ and $b_2$ can be either positive or negative, because the cross-product word-of-mouth can be either positive or negative suggesting a collaboration or a competition between the two products, respectively. Observe that positive or negative cross-product word-of-mouth entirely depends on the sign of $b_1$ and $b_2$ because parameter $\alpha_j,j=1,2$ is always positive. 
\begin{table}
  \centering 
  \begin{tabular}{l|c|l}
\hline
&Product 1  &  Product 2 \\
\hline
  Lotka-Volterra with churn, LVch & $0<\alpha_2<1$ & $0<\alpha_1<1$ \\
  UCRCD & $\alpha_2=1$ & $\alpha_1=1$ \\
  Independent Bass models & $\alpha_2=0$ & $\alpha_1=0$ \\
  Direct cannibalisation &$\alpha_2=0$ & $\alpha_1=1$ \\
  Inverse cannibalisation &$\alpha_2=1$ & $\alpha_1=0$ \\
\hline
\end{tabular}
\caption{Lotka-Volterra with churn model and possible reductions because of parameters $\alpha_1$ and $\alpha_2$}
 \label{tab0}
\end{table}

%Parameters $\alpha_1$ and $\alpha_2$ operate on both the word-of-mouth and the residual market potentials and control a migration --``churn''-- effect between the two competitors, since part of the residual market of one product may migrate to the competitor. \\
Interestingly, the modulation of parameters $\alpha_1$ and $\alpha_2$ allows for the isolation of some specific cases, useful to represent different market environments, as follows:
\begin{enumerate}
\item if $0<\alpha_1<1$ and $0<\alpha_2<1$, we have the full Lotka-Volterra with churn model, where both products are affected by within-product and cross-product word-of-mouth and each one may have access to a portion of the other's residual market;
\item if $\alpha_1=\alpha_2=1$, the LVch model reduces to the UCRCD by \cite{guseomortarino:14}. In this case, the market potential is a common resource $m=m_1+m_2$, and the residual $m-z(t)$, with $z(t)=z_1(t)+z_2(t)$, is completely accessible to both competitors,
\item if $\alpha_1=\alpha_2=0$, there is no competition between the two products, which are in fact described through independent standard Bass models, \cite{bass:69};
\item if $\alpha_1=1$ and $\alpha_2=0$, the life cycle of the first product is described with a standard independent Bass model. The second product's residual market is made by the sum of both residual markets, $(m_2-z_2(t))+(m_1-z_1(t))$ because $\alpha_1=1$. In this sense, the total asymmetry of competition, where the second product has complete access to the residual market of the first, can be seen. At the same time, the first product may still have an impact on the sales of the second by means of the cross-product word-of-mouth, $[{\alpha_1 b_2z_1(t)}/{(m_2+\alpha_1m_1)}]$, which may be either positive or negative depending on the sign of parameter $b_2$. This case illustrates the\textit{ standard product cannibalisation};
\item if $\alpha_1=0$ and $\alpha_2=1$, similar considerations hold. In this case, the first product acts as a winning competitor, cannibalising the market of the second, while the second is described with a standard independent Bass model. This case illustrates a special type of cannibalisation, what we call \textit{inverse product cannibalisation}. 
\end{enumerate}
In Table \ref{tab0}, we summarise all these possibilities depending on the values taken by parameters $\alpha_1$ and $\alpha_2$.

\subsection{Aspects of statistical inference and estimation}
\label{sect:3}
A robust statistical implementation of the models presented in the previous section is based on nonlinear least squares (NLS) under a convenient stacking of the two submodels (see \cite{seber:89}). 
The stacking procedure is suggested  to obtain a unidimensional nonlinear model estimated with standard  NLS methodology under the Levemberg--Marquardt algorithm to overcome some convergence aspects in Newton or quasi-Newton procedures. 

In particular, we consider the structure of a nonlinear regression model:
\begin{equation}
\label{4}
w(t)=\eta(\beta,t)+\varepsilon(t),
\end{equation}
where $w(t)$ is the observed response, $\eta(\beta,t)$ is the deterministic component describing instantaneous or cumulative processes, depending on parameter set $\beta$  and time t, and $\varepsilon(t)$ is a zero mean residual term, not necessarily independent identically distributed (i.i.d.) and normally distributed.

The performance of an extended model, $m_2$, compared with a nested one, $m_1$, may be evaluated through a
squared multiple partial correlation index $\tilde{R}^2$ in the interval $[0;1]$, namely:
\begin{equation}\label{5}
\tilde{R}^2 = (R_{m_2}^2 - R_{m_1}^2)/(1 - R_{m_1}^2),
\end{equation}
where $R_{m_i}^2, \;i=1,2$ are the standard determination indexes of models $m_i, \; i=1,2$.\\
The $\tilde{R}^2$ index has a monotone correspondence with the $F$-ratio:
\begin{equation}\label{6}
F = [\tilde{R}^2 (n-v)]/[(1 - \tilde{R}^2) u],
\end{equation}
where $n$ is the number of observations, $v$ the number of parameters of the extended model $m_2$ and $u$ the incremental number of
parameters from $m_1$ to $m_2$.
Under strong conditions on the distributional shape of the error term $\varepsilon(t)$, especially independence, identical distribution and normality, the $F$-ratio statistic for the null hypothesis 
of equivalence of the two models is a central Snedecor's $F$ with $u$ degrees of freedom for 
numerator and $n-v$ degrees of freedom for denominator $F \sim F_{u,n-v}$, \cite{guidolin:15}.
In more general cases, the F-ratio has a robust distributional behaviour, and a common upper threshold 4, for $u=1$, may be a reference for testing the equivalence of two nested nonlinear models; a lower level, about $2$ for $u > 1$. A more general model, $m_2$, is statistically significant with respect to a nested one, $m_1$, if the F-ratio is much higher than 4.
An overparametrisation of a model $m_2$ may be recognised under a high value of determination index $R^2_{m_2}$ with high instabilities in the asymptotic 95\% confidence limits based on linear approximations. In these situations a sound reduction of the model implies a limited reduction of the determination index $R^2_{m_1}$ compared with $R^2_{m_2}$ through an F-ratio lower than $4$. This contraction is strongly supported by a relevant reduction of the ranges in the asymptotic 95\% confidence limits.

\section{On inverse cannibalisation: competition between Apple's iPhone and iPad}
\label{sect:4}
In this section we analyse the interplay between Apple's iPhone and iPad with the competition models presented in Section \ref{sect:2}. 

We observe some preliminary aspects by inspecting the quarterly actual and smoothed sales data of both products (units sold, in millions) in Fig. \ref{fig1}, as follows
\begin{enumerate}
\item the iPhone entered the market in Q3/2007;
\item the iPad entered the market in Q3/2010 and is characterised by an evident declining trend, having already undertaken the life cycle peak;
\item both products show an evident seasonal component;
\item Apple reports sales data for all its products \textit{without making a distinction between product generations}: the data reflect the fact that the same person may have purchased at different times different generations of the same product, for instance replacing the iPhone 5 with the iPhone 6. Moreover, the same person may own both an iPhone and an iPad, purchased at different times, due to budget constraints;
\item for both products the last data point available refers to Q2/2018. However, we estimated our models leaving out the  Q2/2018 data point, which was used for validation. 
\end{enumerate}

To analyse the presence and nature of competition between the two products, we estimated the full LVch model presented in Equation (\ref{1}). It is worth noting that in this model, we just considered the mean trajectory of sales for a long-term strategic perspective without taking into account the evident seasonal pattern, which is relevant only for a short-term prediction, characterising both products. As performed in subsection \ref{subsect:2}, seasonality can be identified and estimated with SARMAX models once the mean behaviour of the series has been identified and estimated, see for instance \cite{billings:13} and \cite{guidolin:14}.

\begin{table}
\centering
\vspace{-.2cm}
{\footnotesize
\begin{center}
\begin{tabular}{c|c|c|c}
  \hline

   $m_a$ & $p_a$ & $q_a$ & $R^2$\\
  \hline
  120& 0.006& 0.290 & 0.996527\\
  (29)& (0.003)& (0.194)\\
   (209)&(0.009) & (0.385)\\
   \hline
\end{tabular}
\end{center}
} 
\caption{Parameter estimates of a standard Bass model for Apple's iPhone before t=13; Marginal linearised asymptotic $95\%$ confidence limits into brackets. Estimates performed on cumulative data.}
\label{tab01}
\end{table}

In Table \ref{tab01}, we provide the estimates of the LVch model for $t<c_2$ (stand-alone phase), that is, before the iPad launch. Accordingly, sales of the iPhone are modelled with a standard Bass model. 
\begin{table}
\centering
\vspace{-.2cm}
{\footnotesize
\begin{center}
\begin{tabular}{c|c|c|c|c|c}
  \hline
   $p_1$ & $a_1$ & $b_1$ &$\alpha_2$&$m_1$ & $R^2$\\
  \hline
   0.0002 & 0.089  & 0.001  &42.16&8706&0.84989\\
  (-0.008) &(-0.212) &(-0.046) &(-1233)&(-196462)\\
  (0.008) &(0.391) &(0.049) &(1318)&(213876)\\
  \hline
  $p_2$ & $a_2$ &$b_2$&$\alpha_1$&$m_2$& $DW$\\
   \hline
   0.016& 0.184 &-81.49&0.0003&470& 2.014\\
   (-0.032) & (-0.433) &(-1802.74) &(-0.009)&(-820)\\
   (0.065) & (0.803) &(1639.76)&(0.010)&(1760)\\
   \hline
\end{tabular}
\end{center}
} 
\caption{Parameter estimates of the full LVch model. Marginal linearised asymptotic $95\%$ confidence limits into brackets. Estimates performed on instantaneous data. The specific market potentials, $m_1$, for the iPhone and $m_2$, for the iPad, are expressed in millions of units.}
\label{tab02}
\end{table}

The parameters of the competition phase are outlined in Table \ref{tab02}. The determination index $R^2= 0.84989$ (performed on instantaneous data) is very high, but the model looks very unstable, as seen in the confidence intervals (CIs) of the parameters, which suggests a convenient model reduction. The high value taken by $\hat \alpha_2$ and the value close to zero taken by $\hat \alpha_1$ indicate a polarisation of the two parameters, according to case 5) in Section \ref{sect:2}. Following this observation, we performed a model reduction and estimated a Lotka-Volterra model with asymmetric competition, LVac, by setting $\alpha_2=1$ and $\alpha_1=0$.
Moreover, we interpreted the low and unstable estimate of parameter $p_{1}$, $\hat p_{1} = 0.0002$, as a signal of the absence of an innovative component for the iPhone within the competition phase: this appears reasonable because the innovation component for the iPhone is already described by parameter $p_{1a}$ in the stand-alone phase. We highlight that the success of the iPhone heavily relied on word-of-mouth, \cite{peres:10}.

Therefore, we estimated a reduced LVac model with $p_{1}= 0$, namely:

\begin{eqnarray}\label{2}
z'_1(t)&=&\left[\frac{a_1z_1(t)+b_1z_2(t)}{m_1+m_2}\right] \left[(m_1-z_1(t))+(m_2-z_2(t))\right]\\
z'_2(t)&=&\left[p_2+\frac{a_2z_2(t)}{m_2}\right] \left[(m_2-z_2(t))\right]\nonumber.
\end{eqnarray}

\begin{table}
\centering
\vspace{-.2cm}
{\footnotesize
\begin{center}
\begin{tabular}{c|c|c|c}
  \hline
  $a_1$ & $b_1$ & $m_1$&$R^2$ \\
  \hline
   0.202  & -0.196  &1886 & 0.840867\\
  (0.095) &(-0.467) &(1595)\\
  (0.309) &(0.075) &(2178)\\
  \hline
  $p_2$ &$a_2$&$m_2$& $DW$\\
   \hline
   0.013 &0.122 & 454 &1.99\\
  (-0.001) &(0.039) & (350) \\
   (0.028) &(0.205)&(557) \\
   \hline
\end{tabular}
\end{center}
} 
\caption{Parameter estimates of the LV model with asymmetric competition, LVac, and $p_{1c}=0$. Marginal linearised asymptotic $95\%$ confidence limits into brackets. Estimates performed on instantaneous data. The specific market potentials, $m_1$, for the iPhone and $m_2$, for the iPad, are expressed in millions of units.}
\label{tab03}
\end{table}

The estimated parameters of the LVac model with $p_{1}= 0$ are outlined in Table \ref{tab03}. 
To understand if this model reduction is plausible, we compared its performance with that of the general LVch model through the tests for nested models presented in Section \ref{sect:3}. 
Based on this, we accepted the reduced model because
$\tilde R^2= (0.84989-0.840867)/(1-0.840867)=0.056$ and $F=0.056(73-10)/(1-0.056)4=0.93$. In other words, the LVch is an over-parametrised model with non-significant components. Notice that for the LVac model, the confidence limits of the estimates in Table \ref{tab03} are much more precise and sound, compared with the corresponding values in Table \ref{tab02}. The non-homogeneous signs of the CIs extremes for $b_1$ and $p_2$ do not imply in this case difficulties in interpretation: the incoherent signs depend on the approximated extremes as being asymmetrically far from the zero value included in the CIs.

The results of the reduced significant model, LVac, show that the residual market for the iPhone is given by $(m_1-z_1(t))+(m_2-z_2(t))$, 
that is, the residual market of the iPad turns out to be completely available to the iPhone. Conversely, for $\alpha_1=0$, 
the residual market for the iPad is just given by $m_2-z_2(t)$, and the cross-product word-of-mouth vanishes, $[{\alpha_1 b_2z_1(t)}/{(m_2+\alpha_1m_1)}]=0$.
We have statistical support for the hypothesis that the iPad is described by an independent standard Bass model and is therefore not influenced by the iPhone. 
The iPhone is affected by the iPad both in negative and positive terms: the iPad implies a positive extension of the iPhone's residual market but also a competitive effect because of a negative cross-product word-of-mouth because parameter $b_1$ is negative, $\hat b_1= -0.196$. 
 
For comparative purposes, we estimated two independent standard Bass models  to describe the life cycle of the iPhone and the iPad separately, thus ignoring any interaction effects between the two products' sales. The results of this application are outlined in Tables \ref{tab04} and \ref{tab05} (notice that these models are estimated on cumulative data and this explains the expected high $R^2$ compared with the determination index in the rate data case). 
\begin{table}
\centering
\vspace{-.2cm}
{\footnotesize
\begin{center}
\begin{tabular}{c|c|c|c}
  \hline
   $m_1$ & $p_1$ & $q_1$ & $R^2$\\
  \hline
  1701& 0.0013& 0.132 & 0.999112\\
  (1626)& (0.0012)& (0.126)\\
   (1776)&(0.0014) & (0.138)\\
   \hline
\end{tabular}
\end{center}
} 
\caption{Parameter estimates of a standard Bass model for Apple's iPhone; marginal linearised asymptotic $95\%$ confidence limits into brackets. Estimates performed on cumulative data. The specific market potential, $m_1$, for the iPhone is expressed in millions of units.}
\label{tab04}
\end{table}

\begin{table}
\centering
\vspace{-.2cm}
{\footnotesize
\begin{center}
\begin{tabular}{c|c|c|c}
  \hline
   $m_2$ & $p_2$ & $q_2$ & $R^2$\\
  \hline
  415& 0.013& 0.143 & 0.997696\\
  (398)& (0.011)& (0.127)\\
   (431)&(0.014) & (0.160)\\
   \hline
\end{tabular}
\end{center}
}
\caption{Parameter estimates of a standard Bass model for Apple's iPad; marginal linearised asymptotic $95\%$ confidence limits into brackets. Estimates performed on cumulative data. The specific market potential, $m_2$, for the iPad is expressed in millions of units.}
\label{tab05}
\end{table}

Interestingly, the market potentials estimated with the Bass model for the iPhone,  $\hat m_1=1701$, and the iPad, $\hat m_2=415$, are smaller than the corresponding potentials obtained with the LVac model, that is, $\hat m_1=1886$ and $\hat m_2=454$. 
\begin{figure}
\begin{center}
\includegraphics[scale=0.14]{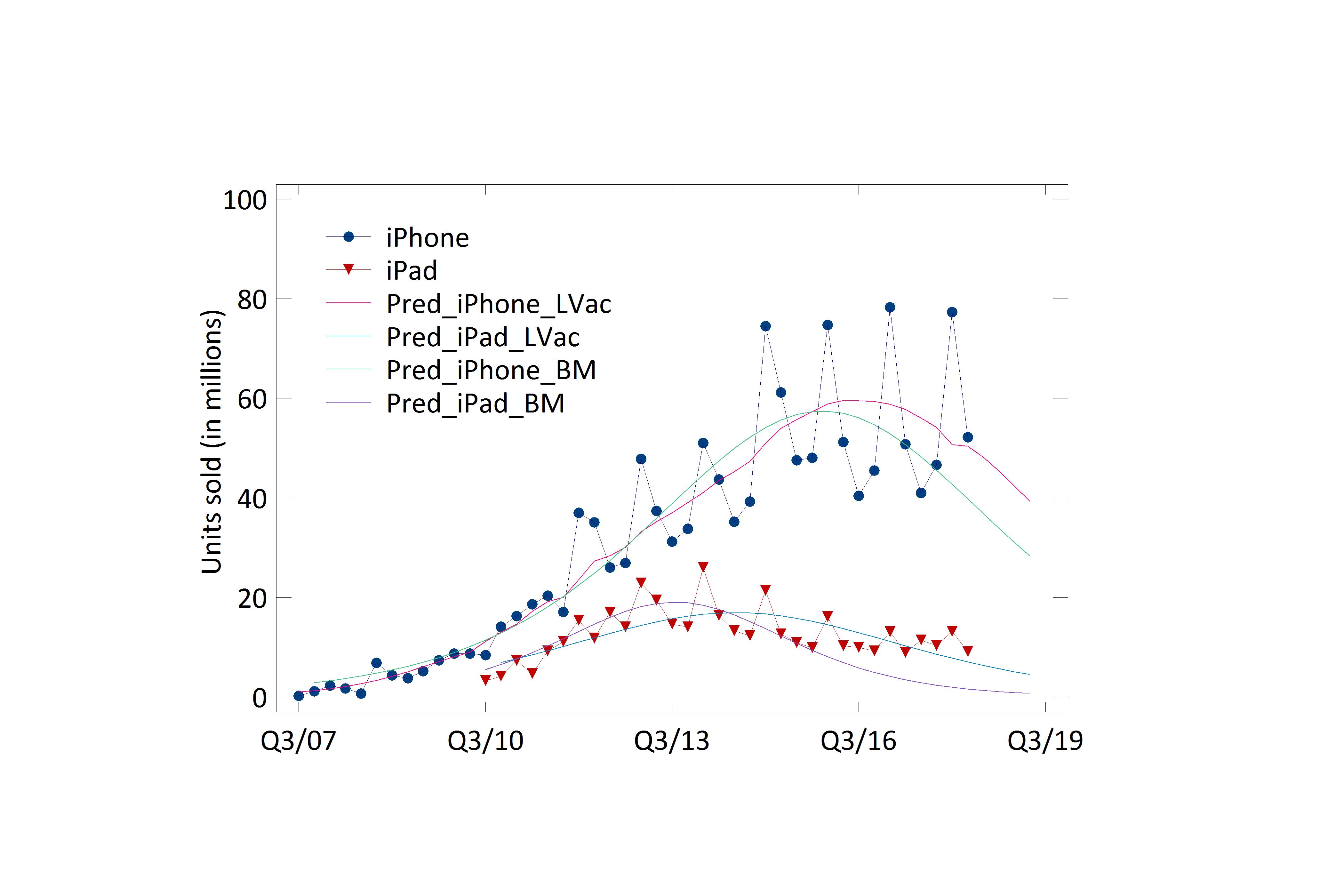}
\caption{Comparison between Lotka-Volterra model with asymmetric competition, LVac,  and Bass model, BM,  for the iPhone and the iPad (data source: Apple Inc.).}
\label{fig2}
\end{center}
\end{figure}

In Fig. \ref{fig2} we perform a graphical comparison between the models: the LVac model can capture  the nonlinear behaviour of the two series, especially in the last part where the predictions are much more relevant. Moreover, Fig. \ref{fig2} suggests that the entrance of the iPad has altered the dynamics of the iPhone's life cycle.

\subsection{Out-of-sample prediction}\label{subsect:2}
The out-of-sample prediction may be performed in two steps:
\begin{enumerate}
  \item {the prediction of nonlinear mean trajectory for each competitor}
  \item {the forecast of the out-of-sample nonlinear time series for each competitor.}
\end{enumerate}

Step $1$ for $z_2(t)$. 

The in--sample and out--of--sample cumulative mean trajectory prediction for $z_2(t)$ is quite simple because we know the exact solution of the second equation in (\ref{2}), namely, a local Bass model, \cite{bass:69}, 

\begin{equation}
W_2(t)- \varepsilon_2(t)=z_2(t)=f_2(\beta,t)= m_2 \frac{1-e^{-(p_2+a_2)t}}{1+(a_2/p_2)e^{-(p_2+a_2)t}}
\end{equation}

where $W_2(t)$ is the observed value and $\varepsilon_2(t)$ is a residual error term. \\
The estimated mean trajectory is, therefore, 

\begin{equation}
\hat z_2(t)= f_2(\hat \beta, t) \;\;\;\;\;\ t=1,2,\dots, T, T+1, \dots,T+K,
\end{equation}
where $\hat \beta$ is the NLS (nonlinear least squares) solution, and $T+1, \dots, T+K$ are the out--of--sample times. 

Step $2$ for $W_2(t)$. 

The in--sample and out--of--sample forecasting of the cumulative observed values may be performed through a SARMAX model (seasonal autoregressive moving average with exogenous input variables), see for instance \cite{billings:13},

\begin{equation}
{_2}\Psi(B^s) {_2}\Phi(B) (W_2(t)-c_2 f_2(\beta,t)) = {_2}\Theta(B){_2}\Omega(B^s)a_t
\end{equation}
where $c_2$ is a real coefficient $\simeq 1$ if $f_2(\beta,t)$ is the correct mean trajectory, ${_2}\Psi(B^s)$, ${_2}\Phi(B)$, ${_2}\Theta(B){_2}$, $\Omega(B^s)$ are backward polynomial operators with specific parameters, $s$ is a seasonal parameter, and ${_2}a(t)$ is a white noise, ${_2}a(t)\sim WN(0,\sigma^2)$. 
SARMAX forecasts, $\bar z_2(t)$,  combine the estimated mean trajectory $f_2(\hat \beta, t)$ and the autodependent effects including seasonal components. 
The residuals  $W_2(t)- \bar z_2(t)= {_2}\hat a(t)$ define an estimated white noise process, $WN(0,  {_2}\hat\sigma^2)$.

Step $1$ for $z_1(t)$. 

The out-of-sample cumulative mean trajectory prediction for $z_1(t)$ depends upon both trajectories $z_1(t)$ and $z_2(t)$. 
Following Euler's approximation $z'_1(t)\simeq z_1(t+1)-z_1(t)$, and denoting by 
\begin{equation*}
c_1(t)= \frac{a_1z_1(t)+b_1z_2(t)}{m_1+m_2}[(m_1-z_1(t))+(m_2-z_2(t))]
\end{equation*}
we obtain a recursive out-of-sample cumulative prediction by exploiting the joint NLS estimates for (\ref{2}), 
\begin{equation}
\tilde z_1(t+1)=z_1(t)+c_1(t), \;\;\;\;\;\;\;\;  t= T, T+1, \dots, T+K-1.
\end{equation}

The estimated mean trajectory in sample is function of the NLS solution, $\hat z_1(t)=\sum_{i=1}^t \hat z_1'(i)$ for $t= 1,2, \dots, T$. \\
The expanded mean trajectory is therefore 
$$\hat z_1(t) =
\bigg \{
\begin{array}{ll}
&\hat z_1(t), \;\;\;t=1,2, \dots, T \\
&\tilde z_1(t), \;\;\; t=T+1,\dots, T+K. \\
\end{array}
$$
Step $2$ for $W_1(t)$. 

The in--sample and out--of--sample forecasting of the observed cumulative values may be performed through a SARMAX model 
\begin{equation}
{_1}\Psi(B^s) {_1}\Phi(B) (W_1(t)-c_1 f_1(\beta,t)) = {_1}\Theta(B){_1}\Omega(B^s)a_t
\end{equation}

SARMAX forecasts, $\bar z_1(t)$, combine the estimated mean trajectory $f_1(\hat \beta, t)$ and the autodependent effects including seasonal components. 
The residuals  $W_1(t)- \bar z_1(t)= {_1}\hat a(t)$ define an estimated white noise process, $WN(0,  {_1}\hat\sigma^2)$. 

\begin{figure}
\begin{center}
\includegraphics[scale=0.14]{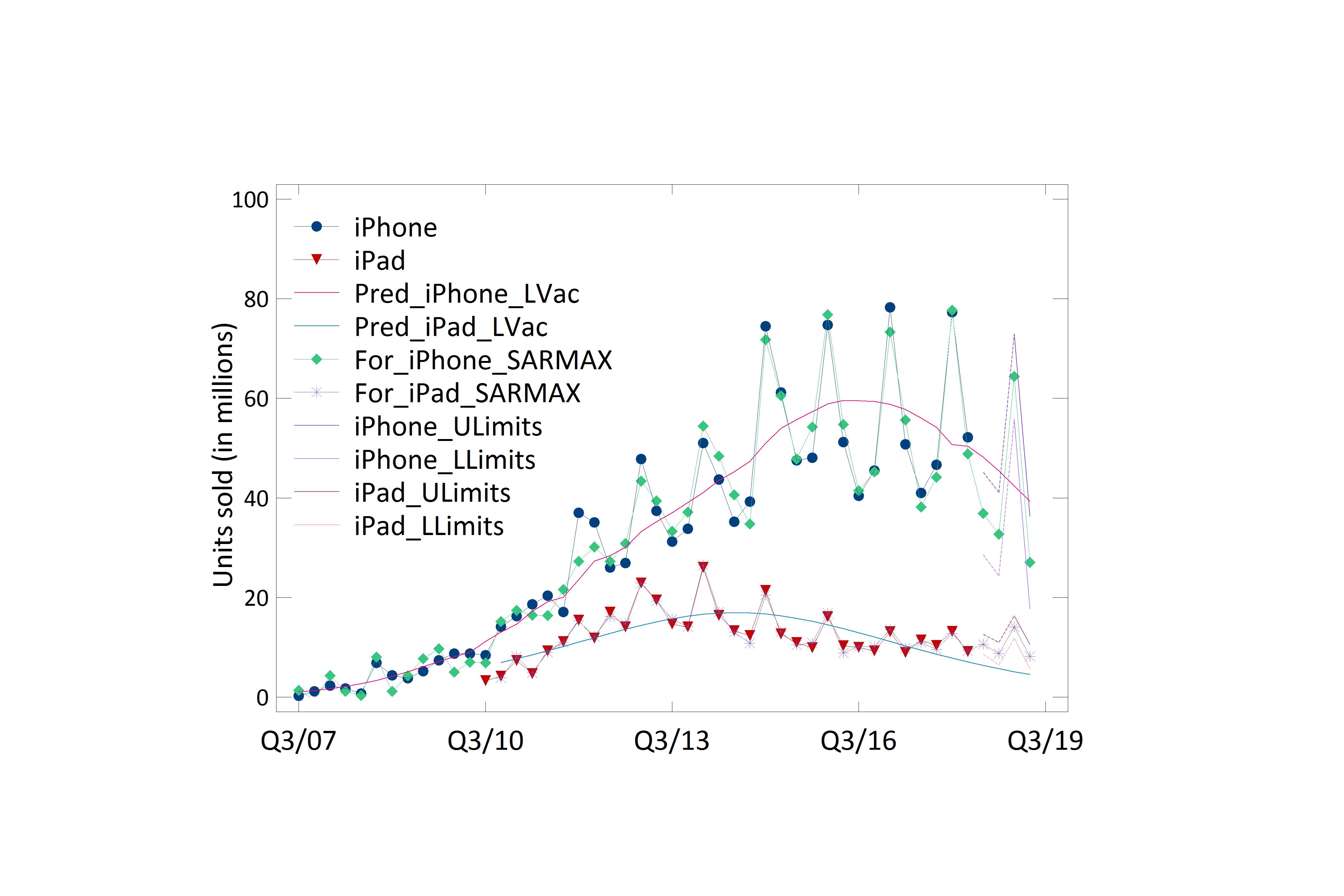}
\caption{In--sample and out--of--sample prediction based on LVac model and SARMAX forecasting for iPhone and iPad. The 95\% limits are provided for out--of--sample uncertainty of forecasts.}
\label{fig3}
\end{center}
\end{figure}

\begin{figure}
\begin{center}
\includegraphics[scale=0.14]{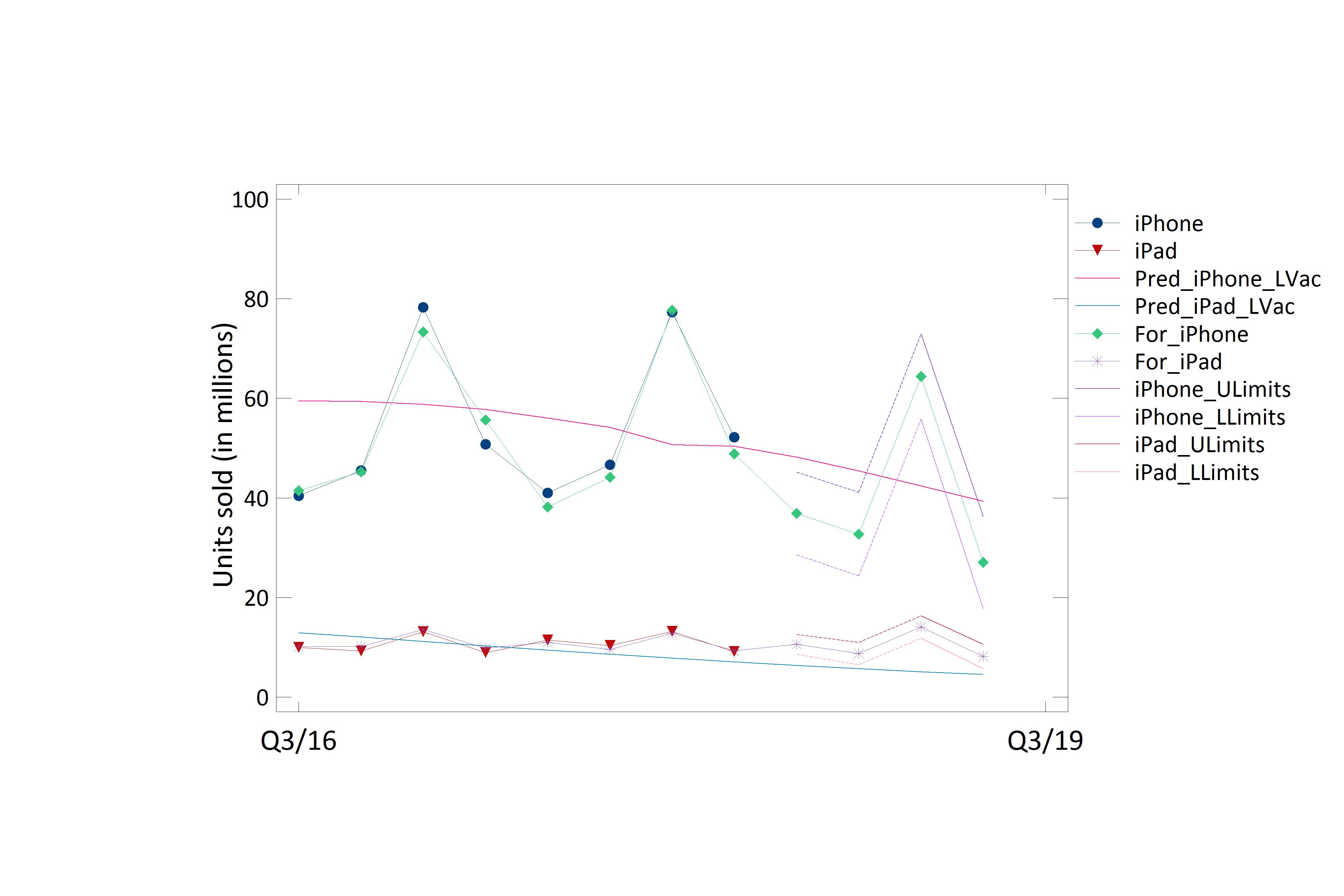}
\caption{A zoom of out--of--sample 95\% forecast limits for iPhone and iPad.}
\label{fig4}
\end{center}
\end{figure}

We may appreciate the performance of the LVac model with SARMAX refinement in Fig. \ref{fig3} and \ref{fig4}, where the 95\% prediction limits are provided for out--of--sample assessment of uncertainty of forecasts.

\section{Lotka-Volterra model with asymmetric competition: \\
a non-dimensional representation}
\label{sect:5}

In this section, we try to evaluate how and to what extent the iPad has altered the life cycle of the iPhone.
To this end, we transform the LVac model in a more tractable non-dimensional representation. \\
Let us reconsider the model selected in Section \ref{sect:4} to treat the case of Apple's iPhone and iPad (for $t > c_2$):

\begin{eqnarray}
z'_1(t)&=&\left[\frac{a_1z_1(t)+b_1z_2(t)}{m_1+m_2}\right] \left[(m_1-z_1(t))+(m_2-z_2(t))\right]\\
z'_2(t)&=&\left[p_2+\frac{a_2z_2(t)}{m_2}\right] \left[(m_2-z_2(t))\right]\nonumber.
\end{eqnarray}

Here, this system depends on six parameters, namely $a_1,b_1,m_1,
p_2,a_2$ and $m_2$. 
This number may be reduced by expressing the system of equations in non-dimensional terms, following, for instance,
\cite{boccara:04}, \cite{laciana:14}.

Non-dimensionalization may recover the characteristic properties of a system of equations  through a convenient scaling of the involved variables. 
To this end, consider the following rescaled variables $x_1=z_1/z_{10}$, $x_2=z_2/z_{20}$, $\tau=t/t_0$.\\
By setting $z_{20}=m_2$ and $z_{10}=m_1$, Equation (\ref{2}) may be rewritten as the following:

\begin{eqnarray}\label{3}
x'_1&=&\left[\frac{a_1z_{10}x_1+b_1z_{20}x_2}{m_1+m_2}\right] \frac{t_0}{z_{10}}\left[(m_1-z_{10}x_1)+(m_2-z_{20}x_2)\right]\\
x'_2&=&\left[p_2\frac{t_0}{z_{20}}+a_2x_2\frac{t_0}{z_{20}}\right] \left[(m_2-z_{20}x_2)\right]\nonumber.
\end{eqnarray}

A further reduction is obtained through $s=m_2/m_1$:

\begin{eqnarray}\label{4}
x'_1&=&\left[\frac{a_1x_1t_0+b_1sx_2t_0}{1+s}\right] \left[(1-x_1)+s(1-x_2)\right]\\
x'_2&=&\left[p_2{t_0}+a_2x_2{t_0}\right] \left[(1-x_2)\right]\nonumber 
\end{eqnarray}

and setting $t_0=1/a_2$ and $r=p_2/a_2$, we obtain the following:

\begin{eqnarray}\label{5}
x'_1&=&\left[\frac{a_1x_1+b_1sx_2}{a_2(1+s)}\right] \left[(1-x_1)+s(1-x_2)\right]\\
x'_2&=&\left(r+x_2\right) \left[(1-x_2)\right]\nonumber.
\end{eqnarray}

The form $(a_1x_1+b_1sx_2)/(a_2(1+s))$ may be multiplied by $a_2(1+s)/b_1$, obtaining $(a_1/b_1)x_1+sx_2$.
Moreover, if we set $v=a_1/b_1$, we obtain a non-dimensional representation of Equation (\ref{2}), which is based on only three parameters:

\begin{eqnarray}\label{6}
x'_1&=&(vx_1+sx_2)\left[(1-x_1)+s(1-x_2)\right]\\
x'_2&=&\left(r+x_2\right) \left[(1-x_2)\right]\nonumber.
\end{eqnarray}

\subsection {Peak conditions for $x'_1$}

Let us indicate with $F_2= (1-e^{-(r+1)\tau})/(1+1/re^{(r+1)\tau})$ the solution of the second equation, a Riccati equation, in (\ref{6}) and rewrite the first equation accordingly:
\begin{equation}\label{7}
x'_1=(vx_1+sF_2)(-x_1+1+s(1-F_2)).
\end{equation}
Taking the first derivative of $x'_1$ with respect to $x_1$ and setting it equal to zero, we obtain the maximum density condition, $\hat x_1$, as follows:
\begin{equation}\label{8}
\hat x_1=\frac{1}{2}+\frac{s}{2}(1-F_2)-\frac{s}{2v}F_2= \frac{1}{2}+\frac{s}{2}\left(1-F_2-\frac{F_2}{v}\right).
\end{equation}

Because $v=a_1/b_1$ is typically negative because $b_1$, which is expressing the cross word-of-mouth effect, is negative, $(1-F_2-F_2/v)$ will be positive. \\
Noting that $s=m_2/m_1$, we may rewrite $\hat z_1=m_1\hat x_1$ in a more interesting form:

\begin{equation}\label{9}
\hat z_1=m_1 \hat x_1=\frac{m_1}{2}+\frac{m_2}{2}\left(1-F_2-\frac{F_2}{v}\right).
\end{equation}

Equation (\ref{9}) highlights that as long as the market potential of the second entrant $m_2$ increases, the maximum peak for the first product, here indicated as $\hat{z}'_1=m_1\hat{x}'_1$, is delayed and reached for $\hat z_1$ beyond $m_1/2$.\\
Following Equation (\ref{8}), we  have that the maximum peak for the iPhone, is reached in the normalized form, at

\begin{equation}\label{10}
\hat x_1=\frac{1}{2}+0.12036(1+0.030612 \;F_2(\tau)).
\end{equation}
It is important to observe that the maximum peak depends on the ratio of the two market potentials, $m_2/m_1$, and on the ratio of the within and cross word-of-mouth effects for the first product (iPhone), $v=a_1/b_1$. \\
Because $F_2(\tau)$ is a cumulative distribution function that may take values between $0$ and $1$, we  have that the maximum peak, $\hat x'_1$ is reached within the interval $0.6204 \leqslant \hat x_1 \leqslant 0.6240$.
Noting that in a standard  life cycle model, such as the Bass model, the peak is reached when $\hat x_1=0.5$, this result indicates that the peak of sales for the iPhone is reached later, and its life cycle is extended thanks to the presence of the iPad.

\section{Summary of results, discussion and concluding remarks}
\label{sect:6}
Information and communication technology markets are characterised by growth strategies that may generate sales cannibalisation. As highlighted in \cite{novelli:13}, the occurrence of cannibalisation must be \textit{tested} and \textit{quantified}, even though this could be a challenging task in ICT markets because the dynamics of technological innovation in ICTs may alter, hide or confound cannibalisation effects.
In this paper we  proposed a competition model, the Lotka-Volterra with asymmetric competition, LVac,  to test statistically the existence and the extent of a special case of product cannibalisation, to our knowledge not modelled in literature, which we call \textit{inverse product cannibalisation}, to express the idea that it is the first entrant that steals market shares from the second. The LVac model has been obtained with a model reduction as a special case of the LVch by \cite{guidolin:15}. 
A convenient model reduction procedure based on the F-ratio test has proven essential for model selection to detect an otherwise hidden market mechanism. 

Through the LVac, we studied the interaction between two products by Apple Inc., finding that the life cycle of the iPad may be described with a simple Bass model, while the iPhone's life cycle has a more complex structure that accounts for the influencing presence of the iPad. 
The iPad had both a negative and positive role: on the one hand, the iPad has exerted competition on the iPhone through a negative word-of-mouth, balancing the specific within-product internal effect with contrasting alternative information, but its presence has also been beneficial because its residual market potential has been completely available to the iPhone, pointing to a final absorption of $m_2$. 
Moreover, a non-dimensional representation of the proposed LVac model evidenced that the entrance of the iPad implied a delay in the peak time of the iPhone: in particular, we found that this delay is because of the size of the market potentials, $m_1$ and $m_2$, and the word-of-mouth dynamics of the iPhone, namely within-product word-of-mouth $a_1$ and cross-product word-of-mouth $b_1$. 
From a strategic perspective, this finding confirms the complex dynamics underlying word-of-mouth \cite{thiriot:18} and points out that even a negative, though limited, spread of information can contribute to determine a product's success. 

The phenomenon of inverse cannibalisation, diagnosed by the LVac model in the iPhone and iPad case through a specific reduction of a more general model, LVch,  suggests similar applications in other contexts, such as those mentioned in the introduction. These may help generalising the proposed concept of inverse cannibalisation. 
There are at least one extension of the current version of the LVac model: 
the inclusion of external control functions mimicking the structure of the generalised Bass model, GBM, \cite{bass:94}, to introduce covariate dependence effects due to external shocks. 
As proposed in Section \ref{sect:4}, the application of standard separate univariate diffusion of innovations models erroneously cancel out the interaction effects that are recovered through a parsimonious model, LVac, allowing a sound interpretation of the process. In particular, its non-dimensional analysis confirms the beneficial effects on the iPhone in terms of the asymptotic market size and the life cycle duration.


\begin{thebibliography}{99}

\bibitem{abramson:98}
G. Abramson, D.H. Zanette,
\newblock Statistics of extinction and survival in Lotka-Volterra systems,
\newblock \emph{Phys. Rev. E\/} \textbf{57}(4) (1998) 4572 4577.


\bibitem{balaz:12}
V. Bal\'{a}\v{z}, A.M. Williams,
\newblock Diffusion and competition of voice communication technologies in the Czech and Slovak Republics, 1948--2009,
\newblock \emph{Technol. Forecast. Soc. Chang.\/} \textbf{79}(2) (2012) 393 404.


\bibitem{bass:69}
F.M. Bass, 
A new product growth model for consumer durables, 
\textit{Manage. Sci.} \textbf{15} (1969) 215 227.


\bibitem{bass:94}
F.M. Bass, T.V. Krishnan, D.C. Jain, 
Why the Bass model fits without decision variables,
\textit{Market. Sci.} \textbf{13}(3) (1994) 203 223.


\bibitem{billings:13}
S.A. Billings, 
Nonlinear system identification: NARMAX methods in time, frequency, and spatio-temporal domains, 
Wiley New York (2013)


\bibitem{boccara:04}
N. Boccara,
\newblock Modelling Complex Systems, 
\newblock Springer New York (2004). 


\bibitem{bordley:17}
R. Bordley, A. Karnani,
Using incentives to address cannibalization, 
\textit{{Long Range Plann.}} \textbf{1} 8 (2017) in press.

\bibitem{jensen:11}
S. Bornholdt, M. H. Jensen, K. Sneppen,
Emergence and Decline of Scientific Paradigms, 
\textit{Phys. Rev. Lett.} \textbf{106} (2011) 058701. 



\bibitem{chakrabarti:16}
A.S. Chakrabarti, 
Stochastic Lotka--Volterra equations: A model of lagged diffusion of technology in an interconnected world, 
\textit{Physica A} \textbf{442} (2016) 214 223. 



\bibitem{chakrabarti:16b}
A.S. Chakrabarti, S. Sinha,
``Hits'' emerge through self-organized coordination in collective response of free agents,
\textit{Phys. Rev. E} \textbf{94} (2016) 042302.  


\bibitem{copulsky:76}
W. Copulsky,
\newblock{Cannibalism in the marketplace},
\textit{J. Marketing} \textbf{2} (1976) 103 105. 

\bibitem {davis:06}
P. Davis,
Measuring the business stealing, cannibalization and market expansion effects of entry in the US motion picture exhibition market, 
\textit{J. Ind. Econ.} \textbf{54} (2006) 293 321. 

\bibitem{guidolin:14}
M. Guidolin, R. Guseo,
Modelling Seasonality in Innovation Diffusion,
\newblock \emph{Technol. Forecast. Soc. Chang.\/} \textbf{86} (2014) 33 40. 


\bibitem{guidolin:15}
M. Guidolin, R. Guseo,
Technological change in the U.S. music industry: within-product, cross-product and churn effects between competing blockbusters, 
\emph{Technol. Forecast. Soc. Chang.\/} \textbf{99 }(2015) 35 46.


\bibitem{gupta:16}
R. Gupta, K. Jain, 
Competition effect of a new mobile technology on an incumbent technology: An Indian case study,
\textit{Telecommun. Policy} \textbf{40} (2016) 332 342. 

\bibitem{guseomortarino:10}
R. Guseo, C. Mortarino,
\newblock Correction to the paper ``Optimal product launch times in a duopoly: balancing life-cycle revenues with product cost'',
\newblock \textit{Oper. Res.\/} \textbf{{58}}(5) (2010) 1522 1523.

\bibitem{guseomortarino:12}
R. Guseo, C. Mortarino,
\newblock Sequential market entries and  competition modelling in multi-innovation diffusions,
\newblock \textit{Eur. J. Oper. Res.\/} \textbf{{216}}(3) (2012) 658 667.

\bibitem{guseomortarino:14}
R. Guseo, C. Mortarino,
\newblock Within-brand and cross-brand word-of-mouth for sequential multi-innovation diffusions,
\newblock \emph{IMA J. Man. Math.\/} \textbf{25}(3) (2014) 287 311.

\bibitem{guseomortarino:15}
R. Guseo, C. Mortarino,
 Modeling competition between two pharmaceutical drugs using innovation diffusion models, 
 \textit{Ann. Appl. Stat.}, \textbf{9}(4) (2015) 2073 2089.


\bibitem{krishnan:2000}
T.V. Krishnan, F.M. Bass, V. Kumar,
\newblock {Impact of a late entrant on the diffusion of a new product/service},
\newblock {\emph{J. Marketing Res.}} \textbf{XXXVII} (2000) 269 278.

\bibitem{laciana:14}
C.E. Laciana, G. Gual, D. Kalmus, N. Oteiza-Aguirre, S.L. Rovere, 
Diffusion of two brands in competition: Cross-brand effect, 
\textit{Physica A} \textbf{413} (2014) 104 115. 


%\bibitem{oteiza:14}
%C.E. Laciana, N. Oteiza-Aguirre, 
%An agent based multi-optional model for the diffusion of innovations, 
%\textit{Physica A} \textbf{394} (2014) 254 265.

\bibitem{meade:06}
N. Meade, T. Islam,
\newblock {Modelling and forecasting the diffusion of innovation - a 25-year review},
\newblock {\textit{Int. J. Forecasting}} \textbf{{22}}(3) (2006) 519 545.


\bibitem{morris:03}
S.A. Morris, D. Pratt,
\newblock Analysis of Lotka-Volterra competition equations as a technological subsitution model,
\newblock \emph{Technol. Forecast. Soc. Chang.\/} \textbf{70}(2) (2003) 103 133.


\bibitem{nortonbass:87}
J.A. Norton, F.M. Bass,
A diffusion theory model of adoption and substitution for successive generations of high-technology products, 
\textit{Manage. Sci.} \textbf{33}(9) (1987) 1069 1086.


\bibitem{nortonbass:92}
J.A. Norton, F.M. Bass,
Evolution of technological generations: the law of capture, 
\textit{Sloan Manage. Rev.} (1992) \textbf{33}(2), 66 77.


\bibitem{novelli:13}
F. Novelli,  
Measuring Sales Cannibalization in Information Technology Markets: Conceptual Foundations and Research Issues, 
In: Herzwurm G., Margaria T. (eds) Software Business. From Physical Products to Software Services and Solutions. ICSOB 2013. 
Lecture Notes in Business Information Processing, vol 150. Springer, Berlin, Heidelberg


\bibitem{peres:10}
R. Peres, E. Muller, V. Mahajan,
\newblock {Innovation diffusion and new product growth models: A critical review and research directions},
\newblock {\emph{Intern. J. Res. Mark.}} \textbf{27}(2) (2010) 91 106.


\bibitem{savin:05}
 S. Savin, C. Terwiesch,
\newblock {Optimal product launch times in a duopoly: balancing life-cycle revenues with product cost},
\newblock {\textit{Oper. Res.\/}} \textbf{{53}}(1) (2005) 26 47.


\bibitem {seber:89}
G.A.F. Seber, C.J. Wild,
\newblock Nonlinear regression,
\newblock Wiley New York (1989).


\bibitem{sornette:04}
D. Sornette, F. Desch\^{a}tres, T. Gilbert, and Y. Ageon, 
Endogenous Versus Exogenous Shocks in Complex Networks: An Empirical Test Using Book Sale Rankings,
\textit{Phys. Rev. Lett.} \textbf{93} (2004) 228701. 



\bibitem{srinivasan:05}
S. R. Srinivasan, S. Ramakrishnan S.E. Grasman,
\newblock Incorporating cannibalization models into demand forecasting,
\newblock \textit{Mark. Intell. Plann.} \textbf{23}(5) (2005) 470 485. 

\bibitem{thiriot:18}
S. Thiriot, 
Word-of-mouth dynamics with information seeking: Information is not (only) epidemics, 
\textit{Physica A} \textbf{492} (2018) 418 430.


\bibitem{thota:11}
H. Thota, Z. Munir, 
Key concepts in innovation, Palgrave Macmillan Los Altos (2011).

\bibitem{tseng:14}
F.M. Tseng, Y.L. Liu, H.H. Wu, Market penetration among competitive innovation products: The case of the Smartphone Operating System,
\textit{J. Eng. Technol. Manage.} \textbf{32} (2014) 40 59. 

\bibitem{vanheerde:10}
H. Van Heerde, S. Srinivasan, M. Dekimpe,
Estimating cannibalization rates for pioneering innovations,
\textit{Market. Sci.} 29 (6) (2010) 1024 1039. 


\bibitem{vitanov:10}
N.K. Vitanov, Z.I. Dimitrova, M. Ausloos, 
Verhulst--Lotka-Volterra (VLV) model of ideological struggle, 
\textit{Physica A} \textbf{389} (2010) 4970 4980. 


\end{thebibliography}
\end{document}